\documentclass[prb,preprint]{revtex4-1} 

\usepackage{amsmath}  
\usepackage{amsfonts} 
\usepackage{graphicx} 

\begin{document}


 \title{Unitary approach to the quantum forced harmonic oscillator}


 \author{
  D. Velasco-Mart\'inez$^1$,
  V. G. Ibarra-Sierra$^2$,
  J. C. Sandoval-Santana$^3$,
  J.L. Cardoso$^1$
  and
  A. Kunold$^1$
 }
 \affiliation{ 
  $^1$ \'Area de F\'isica Te\'orica y Materia Condensada,
   Universidad Aut\'onoma Metropolitana at Azcapotzalco,
   Av. San Pablo 180, Col. Reynosa-Tamaulipas, Azcapotzalco,
   02200 M\'exico D.F., M\'exico 
   \\
   $^2$ Departamento de F\'isica, Universidad Aut\'onoma Metropolitana at
   Iztapalapa, Av. San Rafael Atlixco 186, Col. Vicentina,
   09340 M\'exico D.F., Mexico
   \\
   $^3$  Instituto de F\'isica, Universidad Nacional Aut\'onoma
   de M\'exico, Apartado Postal 20-364,
   M\'exico Distrito Federal 01000, Mexico
   } 
 

 \date{\today}

 \begin{abstract}
In this paper we introduce an alternative approach to studying the evolution
of a quantum harmonic oscillator subject to an arbitrary time dependent force.
With the purpose of finding the evolution operator, certain unitary 
transformations are applied successively to Schr\"odinger's equation reducing 
it to its simplest form. Therefore, instead of solving the original 
Schr\"odinger's partial differential equation in time and space the problem is 
replaced by a system of ordinary differential equations. From the obtained 
evolution operator we workout the propagator. Even though we illustrate the 
use of unitary transformations on the solution of a forced harmonic 
oscillator, the method presented here might be used to solve more complex 
systems. The present work addresses many aspects regarding unitary 
transformations and the dynamics of a forced quantum harmonic oscillator that 
should be useful for students and tutors of the quantum mechanics courses at 
the senior undergraduate and graduate level.
 \end{abstract}


 \maketitle 

 \section{Introduction} \label{Introduction}

The analytic solution of the one-dimensional harmonic oscillator's
Schr\"odinger's equation is one of the first triumphs of the undergraduate
student in the quantum mechanics course. Usually, in a first approach, the 
harmonic oscillator's (HO) energy eigenfunctions and eigenstates are obtained
by solving Schr\"odinger's differential equation. These are recalculated later
by defining the ladder operators. Furthermore, the harmonic oscillator is a 
problem with a wide range of applications in modern physics. It is of enormous 
practical importance in quantum optics and solid state physics for example.
Many of these problems involve the application of an external force to the 
harmonic oscillator. Of special interest is a bounded electrical charge 
subject to an oscillating  electromagnetic field that under certain 
approximations can be reduced to a forced harmonic oscillator.
 
The quantum forced harmonic oscillator (QFHO) has been treated either by 
solving the corresponding differential Schr\"odinger's equation or by 
computing Feynman's path integrals. In the  first case the solution to 
Schr\"odinger equation might be worked out through a series of standard 
variables  changes, \cite{Popov1970} by proposing Gaussian wave function 
\cite{Husimi1953} or by defining a rather general form of the well known 
ladder operators. \cite{Kim1996} Any of these procedures yields the QFHO's 
evolution operator and propagator. Feynman's path integral method, on the 
other hand, \cite{Khandekar1978, Feynman1948, Feynman1950, Merzbarcher3th} 
involves the calculation of the classical action  in order to obtain the 
quantum propagator. The QFHO quantum propagator posses a structure similar to 
the well known propagator for the simple quantum oscillator plus an 
interaction-dependent correction due to the forcing term in the Hamiltonian.
\cite{Merzbarcher3th,Schwinger1951} 

In this paper we seek for an alternative procedure to obtain the QFHO's 
evolution operator and propagator. Our approach exploits certain unitary 
transformations \cite{Maamache1996, Fan1990, Maamache1998, Kunold2013} that 
reduce the Floquet operator to its trivial form i.e. the energy operator. This 
method may be applied to a wide variety of systems whose Hamiltonians are more 
involved and have more dimensions than the one presented in this work. We show 
that the evolution operator and propagator can be obtained from the set of 
unitary transformations mentioned above. Finally, from the evolution operator,
we workout the position and momentum operators in the Heisenberg picture in 
terms of the corresponding operators in the Schr\"odinger picture. This 
transformation posses a symplectic form.

This work is organized as follows. In section \ref{solution} we introduce the
philosophy behind the method to reduce Schr\"odinger's equation and compute 
the evolution operator and propagator. The unitary transformations used to 
calculate de evolution operator are listed in section \ref{evolution}. We also
show how the Schr\"odinger's equation is reduced through these transformations.
Additionally, the Heisenberg picture position and momentum operators are
explicitly calculated and it is demonstrated that their relation with the 
Schr\"odinger picture ones is symplectic. The propagator or Green's function is 
calculated in section \ref{green}. Finally in section \ref{conclusions} we 
summarize.

\section{Unitary transformations}\label{solution}

Unitary transformations are very useful tools in quantum mechanics. In 
particular, in the form of an evolution operator, they may describe the 
dynamics of a closed quantum system since they preserve the probability of 
quantum states.

Whereas, any time independent Hamiltonians $\hat H$ indistinctly yields the
very well known evolution operator
\begin{equation}
U=\exp\left(-i\hat Ht/\hbar\right),\label{simplest}
\end{equation}
time dependent ones may lead to two radically different situations. On one 
hand, if the Hamiltonian commutes with it self at any two different times i. e.
$\left[\hat H\left(t_1\right),H\left(t_2\right)\right]=0$ for $t_1\ne t_2$,
the evolution operator is given by $U=\exp\left(-i/\hbar \int_{t_0}^{t}\hat 
H\left(s\right)ds\right)$. On the other, if $\hat H\left(t\right)$ does not 
commute with it self at two different times, the evolution operator is
\begin{equation}
\label{evop} U=\mathcal{T}\exp\left[-i/\hbar \int_{t_0}^{t}\hat H \left( 
s\right)ds\right] 
\end{equation}
where $\mathcal{T}$ is the time ordering operator. Although very often
the evolution operator is hard to find, and for most physically interesting 
potentials only perturbative solutions can be obtained for Hamiltonians in 
this category, some systems allow analytic solutions to be found. Such is 
the case of the one dimensional harmonic oscillator with time dependent 
coefficients, \cite{Maamache1996} and nonlinear Hamiltonian systems. 
\cite{Maamache1998} Although even in the case in which the system falls in the 
first category, the Hamiltonian is too complex to be treated through
the evolution operator in (\ref{simplest}). This is the case of a system
of coupled harmonic oscillators\cite{Fan1990}, the Hamiltonian
does not depende explicitly on time, however its structure is so involved
that it has been studied via unitary tranformation.

The QFHO's Hamiltonian is given by 
\begin{equation}\label{qham}
\hat{H} = \frac{1}{2m} \hat{p}^2 +\frac{1}{2} m\omega^{2} \hat{q}^{2} -f\hat{q},
\end{equation}
where, $\hat{p}_t$ is the energy operator $i\hbar\partial_t$ and  $\hat{q}$, 
$\hat{p}$ are the space and momentum operators while $f\equiv f\left(t\right)$ 
is the  time dependent force. Clearly, the commutator  $\left[\hat H \left( 
t_1\right), \hat H \left(t_2\right) \right] =\hat p\left[f\left(t_2\right)-
f\left(t_1\right)\right]/m$  does not vanish unless $f$ is a constant, 
therefore the QFHO falls into the third category. Thus, the QFHO's evolution 
operator takes the form (\ref{evop}), nevertheless here we propose  the 
following alternative procedure to compute it. First, we consider 
Schr\"odinger's equation
\begin{equation}\label{ec.schrodinger}
\hat{H} \left\vert \psi \left( t\right) \right\rangle =\hat{p}_t \left\vert
\psi\left(t\right)\right\rangle,
\end{equation}
and conveniently introduce the Floquet operator \cite{Heinzpeter2006}
\begin{equation}\label{floquet}
\hat{\mathcal H} =\hat{H} -\hat{p}_t = \frac{1}{2m}\hat{p}^2 +\frac{1}{2}m 
\omega^{2}\hat{q}^{2}-f\hat{q}-\hat{p}_t,
\end{equation}
that allows to write Schr\"odinger's equation in the compact form  
\begin{equation}\label{shro}
\hat{\mathcal H} \left\vert \psi \left(t \right) \right\rangle= 0.
\end{equation}
Lets now assume that there is a set $\left\{ \hat{\mathcal U}_{1}, 
\hat{\mathcal U}_{2},\dots \hat{\mathcal U}_{n}\right\}$ of unitary 
transformations such that if we apply $\hat{U}=\hat{\mathcal U}_{n} ... 
\hat{\mathcal U}_{2}\hat{\mathcal U}_{1}$ to the Floquet operator, it reduces 
to the energy operator as
\begin{equation}\label{unitary}
\hat{U}\hat{\mathcal H}\hat{U}^{\dagger}= -\hat{p}_t,
\end{equation}
removing the Hamiltonian part. If such a transformation does exists, the 
Schr\"odinger's equation takes the form
\begin{equation}
\hat{U} \hat{\mathcal H} \hat{U}^{\dagger} \hat{U} \left\vert \psi \left( 
t\right)\right\rangle = -\hat{p}_t \left(\hat{U} \left\vert \psi \left( 
t\right)\right\rangle\right)=0.
\end{equation}
Reminding that $\hat p_t$ is $\hbar$ times a time derivative, it is easy to
see that $\hat{U}\left\vert\psi\left(t\right)\right\rangle$ is a constant ket,
say
\begin{equation}
\hat{U} \left\vert \psi \left( t\right) \right\rangle = \left\vert \psi \left(
0\right) \right\rangle
\end{equation}
Therefore, if the set of unitary transformations is known, the evolution of a 
quantum state $\psi$ can be easily calculated by multiplying the previous 
equation by the inverse of $U$ ($U^{-1}=U^\dag$) 
\begin{equation}
\left\vert \psi \left( t\right) \right\rangle = \hat{U}^{\dagger} \left\vert 
\psi \left( 0\right) \right\rangle.
\end{equation}
This equation states that $\hat{U}^{\dagger}$ is the time evolution operator.

The Green's function, or the propagator, is calculated as usual in terms of 
the evolution operator as
\begin {equation}\label{greenf}
G \left( q,q^\prime; t,0 \right) =\left\langle q\left\vert U^{\dagger} 
\right\vert q^\prime\right\rangle.
\end{equation}
  
\section{Evolution operator}\label{evolution}

In this section we aim to find a set of unitary operators that comply with 
equation (\ref{unitary}) and an expression for the evolution operator. With 
this objective in mind it is necessary to consider the Hamiltonian symmetry in 
order to propose the unitary transformations that reduce the Floquet operator.
Notice that the QFHO's Hamiltonian is quadratic in position and momentum 
operators, thus our transformations may not contain terms of a superior order.

The purpose of the first transformation is to remove the linear term of the 
position operator $f\hat q$ from equation (\ref{floquet}). We thus apply a 
translation in the position and momentum operators in the form of
\begin{equation} \label{U1}
\hat{\mathcal U}_{1} =\hat{\mathcal U}_{1t} \hat{\mathcal U}_{1q} \hat{\mathcal 
U}_{1p},
\end{equation}
where
\begin{eqnarray}
\hat{\mathcal U}_{1t} &=& \exp\left(\frac{i}{\hbar}S\right),\\
\hat{\mathcal U}_{1q} &=& \exp\left(-\frac{i}{\hbar}\pi\hat{q}\right),\\
\hat{\mathcal U}_{1p} &=& \exp\left(\frac{i}{\hbar} \lambda\hat{p}\right),
\end{eqnarray}
where $S$, $\pi$  and  $\lambda$ are time dependent transformation parameters.
The above transformation yields the following transformation rules on the 
position, momentum and energy operators
\begin{eqnarray}
\hat{\mathcal U}_{1} \hat{q} \hat{\mathcal U}_{1}^{\dag} &=& \hat{q} +\lambda,
\label{trans}\\
\hat{\mathcal U}_{1} \hat{p} \hat{\mathcal U}_{1}^{\dag} &=& \hat p+\pi, 
\label{boost}\\
\hat{\mathcal U}_{1} \hat{p}_{t} \hat{\mathcal U}_{1}^{\dag} &=& \hat{p}_t
+\dot S -\dot\pi\hat{q}+\dot\lambda\hat{p}+\dot\lambda\pi. \label{ener}
\end{eqnarray}
Equation (\ref{trans}) shows that ${\mathcal U}_{1}$ performs a translation
by $\lambda$ to the position operator $\hat q$. Similarly, in (\ref{boost}) 
and (\ref{ener}) reveal that $\hat{\mathcal{U}}_{1}$ also performs shifts in 
momentum and energy. In the particular case where $\pi$ is a constant and 
$\pi=m\dot\lambda$, (\ref{trans}) and (\ref{boost}) characterize a Galilean 
boost. Nevertheless, as we will show below, the most general case where $f$ is 
an arbitrary function of time requires $\lambda$ and $\pi$ to be nontrivial 
functions of time. 
   
Under the action of $\hat{\mathcal U}_{1}$, the Floquet operator 
(\ref{floquet}) becomes
\begin{eqnarray}\label{U1-floquet}
\hat{\mathcal U}_{1} \hat{\mathcal{H}} \hat{\mathcal U}_{1}^{\dag} = 
\frac{1}{2m} \hat{p}^2 +\frac{1}{2}m \omega^{2} \hat{q}^{2} -\hat{p}_t  +\left(
\frac{\pi}{m} -\dot\lambda\right) \hat{p} +\left( m\omega^{2}\lambda -f +
\dot\pi\right)\hat{q} +\mathcal L-\dot S,
\end{eqnarray}
where we have introduced
\begin{equation} \label{lagragian}
\mathcal L = \frac{1}{2m} \pi^2 + \frac{1}{2}m \omega^{2} \lambda^{2} -f 
\lambda-\pi\dot\lambda,
\end{equation}
As mentioned earlier, the purpose of this transformation is to reduce the 
terms proportional to $\hat q$. Nonetheless, due to the action of 
$\mathcal{U}_1$ new terms have appeared that also need to be canceled in order 
to simplify the Hamiltonian. Hence we now proceed to eliminate the terms 
proportional to $\hat q$, $\hat p$ and the independent ones by setting
\begin{eqnarray}
\frac{\pi}{m}-\dot\lambda=0,\label{euler1}\\
-\dot\pi +f-m\omega^2\lambda=0,\label{euler2} \\
\dot S-\mathcal{L}=0.\label{euler3}
\end{eqnarray}
The first of the previous equations is the standard relation between
the classical velocity and momentum $\dot\lambda$ and $\pi$. The second
is Newton's second law of motion for the classical forced harmonic oscillator.
The last equation states the relation between the action $S$ and the Lagrangian
function $\mathcal{L}$. Furthermore, the somewhat astonishing and interesting 
point is that computing the Euler's equations arising from the Lagrangian
$\mathcal{L}$ we obtain
\begin{eqnarray} 
\frac{d}{dt} \frac{\partial \mathcal L}{\partial\dot \pi} -\frac{\partial 
\mathcal L}{\partial\pi}&=&\frac{\pi}{m}-\dot\lambda=0,\\
\frac{d}{dt} \frac{\partial \mathcal L}{\partial\dot \lambda} -\frac{\partial 
\mathcal L}{\partial \lambda}&=&-\left(m\omega^{2}\lambda -f +\dot\pi\right)=0,
\end{eqnarray}
recovering the conditions (\ref{euler1}) and (\ref{euler2}) that make the 
linear terms vanish.

Here, it is important to establish the initial conditions $S \left( t=0 \right)
=0$, $\pi\left(t=0\right)=0$ and $\lambda\left(t=0\right)=0$ on the 
transformation parameters in order to guarantee that $\hat{\mathcal{U}}_{1}$ 
reduces to the identity operator when $t\rightarrow 0$ i.e.
\begin{equation}
U^{\dagger}(t=0)=\hat{\textbf{I}}.
\end{equation}
    
Hence, the $\mathcal{U}_1$ transforms the Floquet operator into
\begin{eqnarray}\label{U1-floquet-reduces}
\hat{\mathcal U}_{1} \hat{\mathcal{H}} \hat{\mathcal U}_{1}^{\dag} = \frac{1}{2m}
\hat{p}^2+\frac{1}{2}m\omega^{2}\hat{q}^{2}-\hat{p}_t,
\end{eqnarray}
the Floquet operator of a simple harmonic oscillator. The first transformation 
has left a time-independent system that falls in the first category of 
dynamics systems that yield the simplest evolution operator (\ref{simplest}) 
as
\begin{equation}\label{simpleU2}
\hat{\mathcal{U}}_2=\exp\left[- \frac{i}{\hbar}\left(\frac{\hat p^2}{2m}
+\frac{1}{2}m\omega^2\hat q^2\right)t\right].
\end{equation}
However, it is instructive to consider the last transformation in the general 
form of a shear in the $\hat q-\hat p$ space
\begin{equation}\label{U2}
   \hat{\mathcal U}_{2}=
   \exp\left[i\frac{\theta}{2\hbar}
   \left(\frac{1}{\Delta}\hat{p}^{2}+\Delta\hat{q}^{2}\right)\right]
\end{equation}
were $\theta$ and $\Delta$ are in general time dependent transformation 
parameters.

The unitary operator $\hat{\mathcal U}_{2}$ acts on position, momentum and 
energy operators as follows
\begin{eqnarray}  
\hat{\mathcal U}_{2} \hat{q} \hat{\mathcal U}_2^\dagger &=& \hat{q}\cos \theta + 
\frac{1}{\Delta} \hat{p}\sin\theta,\\
\hat{\mathcal U}_{2} \hat{p} \hat{\mathcal U}_2^\dagger &=& \hat{p}\cos\theta - 
\Delta \hat{q} \sin \theta,\\
\hat{\mathcal U}_2 \hat{p}_t \hat{\mathcal U}_2^\dagger &=& \hat{p}_t + 
\frac{\dot{\theta}}{2} \left(\frac{1}{\Delta}\hat{p}^2 +\Delta\hat{q}^2 \right),
\end{eqnarray}
where, supposing beforehand that $\Delta$ is a constant, suffices to 
eliminate all the Hamiltonian's terms.

Applying the last  transformation (\ref{U2}) to the transformed Floquet 
operator (\ref{U1-floquet-reduces}) we have
\begin{eqnarray}\label{transfin}
\hat{\mathcal U}_2 \hat{\mathcal U}_1 \hat{\mathcal H} 
\hat{\mathcal U}_1^\dagger \hat{\mathcal U}_2^\dagger
&=&   \left(\frac{\Delta}{m}\cos^2\theta + \frac{m\omega^2}{\Delta} 
\sin^2\theta -\dot{\theta}\right) \frac{\hat{p}^2}{2\Delta}\nonumber\\
&+& \left(\frac{\Delta}{m}\sin^2\theta + \frac{m\omega^2}{\Delta} 
\cos^2\theta-\dot{\theta}\right) \frac{\Delta}{2}\hat{q}^2\nonumber\\
&+& \left(\frac{m\omega^2}{\Delta} - \frac{\Delta}{m}\right) \frac{1}{2} 
\cos\theta\sin\theta \left(\hat{p}\hat{q}+\hat{q}\hat{p}\right) -\hat{p}_t.
\end{eqnarray}
To reduce this last form of the Floquet operator to the energy operator 
$\hat{p}_{t}$ we must eliminate the coefficients of $\hat{q}^2$, $\hat{p}^2$ and 
$\hat p\hat q+\hat q\hat p$ obtaining the system of differential equations
\begin{eqnarray}\label{ecs2}
\frac{\Delta}{m} \cos^2\theta + \frac{m\omega^2}{\Delta} \sin^2\theta 
-\dot{\theta} &=&0,\label{ecs2-1}\\
\frac{\Delta}{m} \sin^2\theta + \frac{m\omega^2}{\Delta} \cos^2\theta 
-\dot{\theta} &=&0,\label{ecs2-2}\\
\frac{m\omega^2}{\Delta}-\frac{\Delta}{m} &=&0\label{ecs2-3}.
\end{eqnarray}
From (\ref{ecs2-3}) we obtain that $\Delta=m\omega$ and consequently 
(\ref{ecs2-1}) and (\ref{ecs2-2}) reduce to 
\begin{equation}
\omega-\dot{\theta}=0.
\end{equation}
Once more, in order for $\mathcal{U}_2$ to reduce to unity, the initial 
condition on the $\theta$ parameter must be set to $\theta(t=0)=0$ giving
\begin{equation}
\theta=\omega t.\label{omegat}
\end{equation}
Notice that the relations above turn the unitary transformation (\ref{U2})
into (\ref{simpleU2}).

In this way all the Hamiltonian terms vanish to yield the simplest Floquet 
operator
\begin{equation}
\hat{\mathcal U}_2 \hat{\mathcal U}_1\hat{\mathcal H}
\hat{\mathcal U}_1^\dagger\hat{\mathcal U}_2^\dagger =-\hat{p}_t.
\end{equation}
As we mentioned above, this is precisely the evolution operator
\begin{equation}
\hat{U}^\dagger=\hat{\mathcal U}_1^\dagger \hat{\mathcal U}_2^\dagger.
\end{equation}
Note that the time evolution operator's time dependence enters through the 
parameters $S$, $\lambda$, $\pi$ and $\theta$. Once the explicit form of the 
force $f$ is given $\pi$ and $\lambda$ can be calculated from the system of 
differential equations formed by (\ref{euler1}), (\ref{euler2}) and the 
initial conditions $\lambda=\pi=0$ with $t=0$. With the expressions of 
$\lambda$ and $\pi$ at hand the Lagrangian $\mathcal{L}$ is a known function 
and $S$ can readily be worked out from (\ref{euler3})
\begin{equation}
S=\int_0^t\mathcal{L}\left(s\right)ds.
\end{equation}
We have therefore reduced the problem to solving a system of
on variable differential equations.
Naturally the difficulty in finding solutions for $\lambda$ and $\pi$
hinges in the complexity of the function $f$.

In the following we obtain the Heisenberg picture position and momentum 
operators' time evolution via the application of the above obtained evolution 
operator
\begin{eqnarray}
   \hat{q}_H(t)&=&\hat{U} \hat{q} \hat{U}^{\dag}
   =\hat{q} \cos \theta+ \hat{p} \frac{1}{\Delta} \sin \theta +\lambda, \\
   \hat{p}_H(t)&=&\hat{U} \hat{p} \hat{U}^{\dag}
   =\hat{p} \cos \theta- \hat{q} \Delta \sin \theta +\pi.
\end{eqnarray}
Rearranging the previous results in in matricial form we get
\begin{equation}
  \begin{bmatrix}
    \hat{q}_H(t)\\
    \hat{p}_H(t)\\
  \end{bmatrix}
  =
  \begin{bmatrix}
    \cos\theta & \frac{1}{\Delta} \sin\theta  \\
      -\Delta\sin\theta & \cos \theta  \\
     \end{bmatrix}
     \begin{bmatrix}
      \hat{q} \\
      \hat{p} 
   \end{bmatrix}
   +
   \begin{bmatrix}
     \lambda \\
      \pi
     \end{bmatrix}
     =
     \mathcal{M} 
     \begin{bmatrix}
      \hat{q} \\
      \hat{p}
     \end{bmatrix}
     +\mathcal{\xi}.
    \end{equation}
where
\begin{equation}
\mathcal{M}=
  \begin{bmatrix}
    \cos\theta & \frac{1}{\Delta} \sin\theta  \\
    -\Delta\sin\theta & \cos \theta  \\
  \end{bmatrix}
=
  \begin{bmatrix}
    \cos\omega t & \frac{1}{m\omega} \sin\omega t  \\
    -m\omega\sin\omega t & \cos \omega t  \\
  \end{bmatrix}
\end{equation}
is clearly a symplectic matrix.

\section{Green's function}\label{green} 
    
Now we focus on obtaining the Green's function using the time evolution 
operator through the relation (\ref{greenf}). The importance of obtaining the 
general form of this function lies in the fact that it provides us with a 
direct method to calculate the evolution of a given wave packet from $t=0$ to 
$t$. 

At this point it is assumed that, given the explicit form of the force $f$,
the solution of equations (\ref{euler1}), (\ref{euler2}) and (\ref{euler3})
for $\lambda$, $\pi$ and consequently $S$ are known. In order to find $G$ we 
insert a complete basis of the position states so we can calculate separately 
the matrix elements of each unitary transformation $\mathcal{U}_1^\dag$ and 
$\mathcal{U}_2^\dag$  as follows
\begin{equation}\label{greenin}
     G(q, q^{\prime};t,0)= \int dq^{\prime\prime} \left\langle q
     \left\vert \mathcal{U}_1^{\dagger}\right\vert q^{\prime 
     \prime}\right\rangle\left\langle q^{\prime \prime} 
     \left\vert \mathcal{U}_2^{\dagger}\right\vert q^\prime\right\rangle.
\end{equation}
The first matrix element, corresponding to the unitary transformation related 
with a translation in position and momentum, is readily find as
 \begin{equation}
     \left\langle q \left\vert \mathcal{U}_1^{\dagger}\right\vert q^{\prime \prime} \right\rangle = 
     \exp\left[-\frac{i}{\hbar}S\right] \exp\left[\frac{i}{\hbar}\pi q^{\prime \prime}\right] \delta(q-q^{\prime \prime} 
     -\lambda).
\end{equation}
The second element corresponding to a shear is
\begin{equation}
     \left\langle q^{\prime \prime} \left\vert \mathcal{U}_2^{\dagger}\right\vert q^\prime\right\rangle =   
     \sqrt{\frac{\Delta}{2 \pi \hbar \sin \theta}} \exp\left[\frac{i \Delta }{2 \hbar \sin \theta} \left[ 
     {(q^{\prime}}^2+{q^{\prime \prime}}^2) \cos \theta - 2 q^{\prime} q^{\prime \prime} \right] \right] .
\end{equation}
Note that this  matrix element is in fact the propagator for a simple harmonic 
oscillator. \cite{Merzbarcher3th} By integrating (\ref{greenin}) we finally 
arrive to the explicit form of the Green's function
\begin{multline}
     G\left(q, q^{\prime};t,0\right)= \sqrt{\frac{\Delta}{2 \pi\hbar\sin\theta}}\exp\left(-\frac{i}{\hbar} S \right)
     \exp\left[\frac{i}{\hbar} \left(\pi -\frac{\Delta}{\hbar \sin\theta} q^{\prime} 
     \right)(q-\lambda)\right] \nonumber\\
     \times\exp\left[\frac{i \Delta}{2 \hbar} \cot\theta( (q-\lambda)^2 +q'^2)\right]\\
     =\sqrt{\frac{m\omega}{2 \pi\hbar\sin\omega t}}\exp\left(-\frac{i}{\hbar} S \right)
     \exp\left[\frac{i}{\hbar} \left(\pi -\frac{m\omega}{\hbar \sin\omega t} q^{\prime} 
     \right)(q-\lambda)\right] \nonumber\\
     \times\exp\left[\frac{i m\omega}{2 \hbar} \cot\omega t( (q-\lambda)^2 +q'^2)\right]
\end{multline}
This function is formed by linear and quadratic terms of the position 
eigenstates. It exhibits the correct structure predicted by Schwinger and 
others \cite{Schwinger1951,Urrutia1984,Pepore2009} for Hamiltonians that are 
quadratic in the  position and momentum operators.

\section{Conclusions} \label{conclusions}
We have obtained the evolution operator and the propagator of a forced 
harmonic oscillator by performing a series of unitary transformations. These 
transformations are chosen to eliminate certain terms of the Floquet operator 
to the point of reducing to its simplest form, the energy operator. In this 
particular case two transformations are needed: the first one shifts position, 
momentum and energy allowing to remove from the Hamiltonian the potential 
energy due to the force. The second and final transformation resembles a shear 
in the $\hat q-\hat p$ space. The first transformation leaves the Floquet 
operator of a simple quantum harmonic oscillator. Finally the shear cancels 
all remaining terms to yield the quantum forced harmonic evolution operator 
and propagator.

The obtained result for the propagator is consistent with the one obtained by 
Feynman integration and with the general structure proposed in 
Ref[\onlinecite{Merzbarcher3th}] and Ref[\onlinecite{Schwinger1951}] .

Despite the simplicity of the forced harmonic oscillator, it may serve to 
illustrate the use of unitary transformations to describe its dynamics. This 
method may be applied to the solution of more complex systems as the one of a
charged particle subject to arbitrary electromagnetic fields. \cite{Kunold2013}
Unfortunately the choice on which unitary transformations should be used to 
reduce the Floquet operator is not systematic nor direct, and should be made 
intuitively for each case.

\acknowledgments
V. G. Ibarra-Sierra would like to thank the scholarship granted by the UAM for 
his doctoral studies.

 
\end{document}